# Predictions Wrong Again on Dead Zone Area - Gulf of Mexico Gaining Resistance to Nutrient Loading

Michael W. Courtney and Joshua M. Courtney
BTG Research, 9574 Simon Lebleu Road, Lake Charles, LA, 70607
Michael_Courtney@alum.mit.edu

**Abstract**
Mississippi River nutrient loads and water stratification on the Louisiana-Texas shelf contribute to an annually recurring, short-lived hypoxic bottom layer in areas of the northern Gulf of Mexico comprising less than 2% of the total Gulf of Mexico bottom area. This paper observes that the NOAA and LUMCON have now published errant predictions of possible record size areas of temporary bottom water hypoxia ("dead zones") three times since 2005, in 2008, 2011, and 2013 and that the LUMCON predictions of the area of hypoxic bottom water average 31% higher than the actual measured hypoxic areas from 2006 to 2014. These systematically high predictions depend on the assumption that the susceptibility of the Gulf of Mexico to forming hypoxic areas in response to nutrient loading has been relatively constant since 2001, though the susceptibility has been occasionally adjusted upward in different models. It has been previously suggested that tropical storms in a given year that occur on the Louisiana-Texas shelf between the peak nutrient loading in spring and formation of the hypoxic zone in summer can mitigate the size of the hypoxic zone that year through mixing of stratified well oxygenated lighter and warmer surface layers and oxygen depleted heavier and cooler bottom layers. This paper suggests several reasons why the Louisiana-Texas shelf may be systematically growing less susceptible to a given level of nutrient loading over time so that predictions based on the measured area of temporary bottom water hypoxia prior to 2006 tend to be too big in recent years.

Measurements of bottom water oxygen levels have documented seasonal hypoxic areas on the Louisiana-Texas shelf almost every year since 1985 (Bianchi et al. 2010, Rabalais et al. 2007, Rabalais et al. 2002). These areas of temporary bottom water hypoxia result from a combination of factors, including nutrient loading from the Mississippi River, allowing a large, seasonal increase in phytoplankton production. In areas where the water is stratified (lighter, warmer, lower salinity waters on top and heavier, cooler, higher salinity waters on the bottom), aerobic bacteria breaking down organic matter can deplete oxygen if sufficient quantities of organic matter reach the bottom as a result from increases in phytoplankton in a short period of time (Bianchi et al. 2010, Rabalais et al. 2007, Rabalais et al. 2002). The time window for possible hypoxia formation tends to be on the order of four to eight weeks long, because the stratification and hypoxia can be broken up by mixing due to passing tropical systems and by cooler weather and wind forced mixing by September.

  Suggestions that these short-lived, relatively small (< 2% of the Gulf of Mexico area), regions of bottom water hypoxia have significant negative impacts on Gulf of Mexico fisheries are misleading and unsupported by data. Grimes (2001) has pointed out that 70-80% of Gulf of Mexico fisheries production comes from the areas surrounding the Mississippi River discharge and that the increased fisheries production results from nutrient-enhanced increases in primary production. Caddy (1993, 2000) points out that fisheries production increases in response to increased nutrient loading up to a point, and then declines. All the available evidence suggests that the Louisiana-Texas shelf is on on the rising edge or the top of fisheries response to nutrient loading and that decreasing nutrient loads delivered by the Mississippi River to the Louisiana-Texas shelf would likely have the effect of decreasing fishery production most years (Courtney and Courtney 2013). It is likely that recent increases in Red Snapper production and total sustainable yield are attributable, in part, to nutrient loading from the Mississippi River discharge (Courtney et al. 2013). Bianchi et al. (2010) point out the apparent paradox of significant increases in faunal biomass in the areas near the Mississippi River plume.

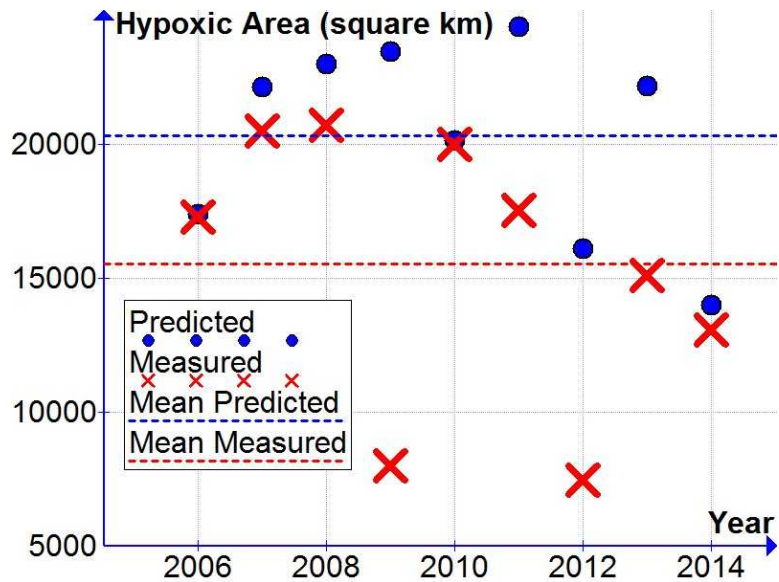

*Figure 1: A comparison of LUMCON's annual predictions for hypoxic zone areas compared with their actual measurements. Their predictions since 2006 are systematically high. This paper suggests reasons why the recalibration of their models is inapplicable after 2005, and that the Gulf of Mexico has actually become more resistant to nutrient loading causing hypoxic zones since 2001. (Data obtained from LUMCON annual predictions and press releases available at www.gulfhypoxia.net ).*

A number of models for predicting the areas of the regions of hypoxic bottom water have been published. Each year there are press releases and other announcements of predictions of the area of bottom water hypoxia expected that summer. These models are based on historical areas of bottom water hypoxia as measured since 1985 (with occasional "recalibrations" suggesting increases in sensitivity) as well as measurements and expectations of the major contributing factors (nutrient discharge, freshwater discharge, salinity, wind forcing, currents, stratification, etc.) The models attempt to account for a number of factors that vary annually, but on the whole, the predictive models assume that the susceptibility of the system for nutrient loading to result in large areas of hypoxic bottom water has not systematically decreased from 1985 to 2013. In other words, given the same inputs, these models would forecast the same or larger areas of hypoxic bottom water in 2013 as they would hindcast for 1985.

The National Oceanic and Atmospheric Administration (NOAA) and LUMCON both published predictions of possible record hypoxic zone areas in 2008, 2011, and 2013. LUMCON also published unfulfilled predictions of record hypoxic areas in 2007 and 2009. These predictions tend to generate more media coverage than the eventual measurement results failing to show record setting "dead zone" areas. (The last record size hypoxic area was in 2002.) NOAA hedges their bet by only predicting "possible" record areas, and LUMCON includes fine print of possible factors that might reduce the hypoxic area. However, neither LUMCON nor NOAA have made efforts to curb the public perception of looming disaster and belief that record "dead zone" areas are portended. In fact, the 2013 NOAA "dead zone" prediction included an errant map showing large red areas purporting to be a region of hypoxic bottom water somehow related to the 2013 prediction (NOAA 2013). In reality, that map was a NASA MODIS/Aqua image showing turbidity from 2004 (NASA 2004) rather than any depiction of bottom water hypoxia (Courtney and Courtney 2013).

A graph of LUMCON's predicted and subsequently measured hypoxic zone areas from 2006 to 2014 is shown in Figure 1. Clearly, the models employed generate predictions that are systematically higher than measured values. The periodic recalibrations of underlying models is only justified by the fact that the recalibrated models fit the data better, and the historical recalibrations suggest increasing sensitivity of the hypoxic area to the nutrient load. To date, the models have not been recalibrated to suggest increasing resistance, nor has there been detailed discussion of physical mechanisms potentially underlying increased resistance.

The purpose of this paper is to suggest that the underlying assumption in the modeling employed in predicting hypoxic zone areas may be in error.  We cannot rule out the possibility that conscious and sub-conscious biases in model choices (inputs and methods) contribute to exaggerated predictions.  Nor can we rule out the possibility that reporting biases tend to exaggerate the size or importance or impact of the seasonal bottom water hypoxia.  Instead, however, we suggest that a variety of factors may be combining to provide the Gulf of Mexico with an increasing resistance to bottom water hypoxia for a given level of nutrient loading from the Mississippi River discharge.  The Louisiana-Texas shelf may be less susceptible to hypoxia resulting from nutrient loading than it was in the 1980s and 1990s and in 2002 when the last record area of bottom water hypoxia was documented.  Any one factor in isolation probably would not have a significant impact on the Gulf of Mexico's response to nutrient loading, but the combined effects of multiple factors may provide a reasonable explanation for the increased resistance inferred from observed hypoxic zone areas systematically lower than model predictions which generally assume a constant susceptibility to nutrient loading since 2001.

The following factors may be contributing to the Gulf of Mexico's growing resistance to hypoxia for a given level of nutrient loading:  1)  Loss of coastal and inshore wetlands and barrier islands has increased the total flux of water and nutrients from the open Gulf of Mexico into inshore estuaries, spreading the nutrients out and decreasing nutrient concentration in the areas most susceptible to stratification and hypoxia.  2)  Larger tidal flows between the Gulf of Mexico and estuaries produce faster currents which cause greater mixing, delaying water stratification and reducing the area of stratified water necessary for formation of hypoxic zones.  3)  The food web has changed over time so that carbon added to the biomass from phytoplankton production at times of peak nutrient availability reaches higher trophic levels (on average), thus remaining in the biomass significantly longer before dying and decaying at the bottom.  Carbon reaching the bottom after the breakup of stratified layers will not contribute to hypoxic areas.  4)  The path in the food web from diatom to copepod feces that results in quick deposition of carbon on the bottom has changed over time in favor of paths where more carbon remains in surface and mid-level waters for longer periods, thus delaying the deposition of carbon in the bottom layers of water.  5)  The average turbidity of the areas of the Louisiana-Texas shelf prone to hypoxia has been increased by silt moved into the Gulf of Mexico by coastal erosion and wetland loss, together with Mississippi River sediments, thus decreasing light penetration and eutrophication.  6)  Debris washed into the Gulf by tropical storms and hurricanes, together with man-made structures (petroleum platforms) increase mixing during strong current flows thus disrupting stratification.  7)  Food webs local to these artificial reefs contribute to more carbon reaching higher trophic levels, delaying the decay of biomass in bottom waters.

LUMCON's *ex post facto* explanations of errant dead zone forecasts tend to emphasize meteorological mixing events, such as tropical systems or atypical wind patterns, as the main reason for the predictions exceeding the measured areas in most years.  We believe that alternate (non-meteorological) explanations need to be considered in light of the number of years the forecasts have been in error.  In either case, the forecasting models seem to be in need of recalibration.  Finally, "recalibrating" the models every few years to better fit the data seems more like tweaking the fudge factor than legitimate science.  Accurate models capable of predicting the susceptibility of the Gulf of Mexico to hypoxia and providing quantitative understanding of the causal factors contributing to that susceptibility would give greater confidence that the government agencies asserting that *nutrients are pollution* are basing that assertion in sound science.

**Acknowledgments**
We thank Amy Courtney, PhD, for helpful discussion and comments that improved the manuscript.

**About the Lead Author**

Michael Courtney's first laboratory job was in Fisheries Science at the LSU Aquaculture facility in 1985-1986. He received the LSU University Medal in 1989 for graduating ranked first in his class with a BS in Physics before pursuing a PhD in Physics from MIT, completing that degree in 1995. He has been an active researcher in blast physics and ballistics since 2001, and served on the Mathematics faculty of the United States Air Force Academy from 2009 to 2013. He returned to research in Fisheries Science while directing the Air Force Academy's Quantitative Reasoning Center and has published numerous papers bringing quantitative analysis and scientific insights to bear on a variety of problems. He currently serves as a consulting scientist for BTG Research (www.btgresearch.org).

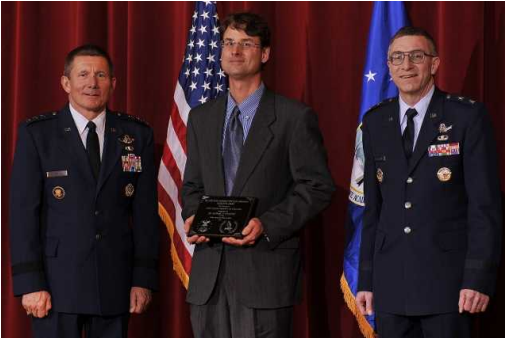

Michael Courtney is shown at left (center) receiving a 2012 research award from Lt. Gen. Mike Gould, superintendent of the U.S. Air Force Academy, and Maj. Gen. William McCasland, commander of the Air Force Research Laboratory, for his research collaborating with cadets.